# An agent-based computational model for China's stock market and stock index futures market

Hai-Chuan Xu[1,2], Wei Zhang[1,2], Xiong Xiong[1,2*], Wei-Xing Zhou[3,4,5]

[1]*College of Management and Economics, Tianjin University, Tianjin 300072, China*
[2]*China Center for Social Computing and Analytics, Tianjin University, Tianjin 300072, China*
[3]*School of Business, East China University of Science and Technology, Shanghai 200237, China*
[4]*Department of Mathematics, East China University of Science and Technology, Shanghai 200237, China*
[5]*Research Center for Econophysics, East China University of Science and Technology, Shanghai 200237, China*

**Abstract:** This study presents an agent-based computational cross-market model for Chinese equity market structure, which includes both stocks and CSI 300 index futures. In this model, we design several stocks and one index futures to simulate this structure. This model allows heterogeneous investors to make investment decisions with restrictions including wealth, market trading mechanism, and risk management. Investors' demands and order submissions are endogenously determined. Our model successfully reproduces several key features of the Chinese financial markets including spot-futures basis distribution, bid-ask spread distribution, volatility clustering and long memory in absolute returns. Our model can be applied in cross-market risk control, market mechanism design and arbitrage strategies analysis.

## 1. Introduction

On April 16, 2010, CSI 300 index futures succeeded in listing on the China Financial Futures Exchange, which are the first index futures in China and provide a new hedge tool for Chinese financial market practitioners. However, the leverage properties of stock index futures, as well as their hedging strategies which may lead to market shocks, will increase the systemic risk, such as South Korea's stock index futures market manipulation in 2006 and Dow Jones Industrial Average's intraday flash crash caused by the E-mini futures contracts. Under the background that serious financial risk events frequently occurred, it is essential to grasp the law of financial risks and then do a good job in risk control and mechanism design, so that we can develop stock index futures and other derivatives successfully. Considering the co-movement between stock index futures market and its underlying stock market, risk management in a single market is not sufficient. It is critically more important for us to focus on and control cross-market risks.

Previous studies show controversial results for the role of index futures on the spot market. Some researchers found that futures market may stabilize the spot market, while others found that index futures can shock the underlying stocks. For examples, Bae et al. and Wang et al. separately analyzed South Korea's KOSPI200 index futures, Hong Kong Hang Seng H index futures and their corresponding spot markets, and suggested that stock index futures increase the volatility of the spot market [1, 2]. Xiong et al. confirmed that derivatives can bring risk to stock market from the view of market manipulation [3]. In contrast, Drimbetas et al. argued that stock index futures reduce the stock volatility based on the analysis on the Greek financial market [4]. These studies show that different results are obtained under different samples and different time intervals. However, due to the limitation of empirical study, the impact mechanisms and transmission channels of cross-market risk are unknown. Recently, scholars have pointed out that the trading mechanisms would affect the role of stock index futures on spot market. Based on 42 stock exchanges, Cumming et al. found that the trading mechanisms have significant influence on market liquidity [5]. Ritter suggested that futures market have spillover effects on the spot market and thus stock index futures take the role of information diffusion to stock market [6]. However, due to the complexity of the market micro-structure and investors' behavior, it is difficult to clearly analyze the interaction effects between the futures market and spot market. Hence it is more difficult to study the cross markets information and risk transmission mechanisms using traditional empirical studies or mathematical methods.

Agent-based computational models, which have developed rapidly in recent years, bring an alternative view angle for the research on the complicated relationships of cross markets. Several famous agent-based computation finance (ACF) labs have done some preliminary studies on the pricing of derivatives and the risk formation using agent-based computational models. Cappellini developed an artificial market called SumWEB where stock market and stock index futures market coexist, which provided a basic platform for studying cross-market arbitrage and price manipulation behaviors [7]. Kobayashi and Hashinoto suggested that the market circuit-breaker would do a good job in controlling excessive price fluctuations and stabilizing the market [8]. Ecca, Marches and Setzu established an agent-based computational model which includes an artificial stock market and an artificial stock options market. They concluded that options would reduce the price volatility of the underlying stock [9].

---

\* Corresponding author: xxpeter@tju.edu.cn (X. Xiong).

However, the models above still stay far away from real market when talking about trading mechanisms or investors' demand decisions. Some scholars even criticized that some of the existing agent-based computational models are toy models. In this direction, Westerhoff established an artificial stock market with heterogeneous interacting agents to consider some supervise mechanisms such as trading taxes, intervention from the central bank and trading suspension. He pointed out that agent-based models would help to understand the role of regulatory policies [10]. Given that it is still a single-market model, it goes no further on the spot-future cross-market interactions.

We design and develop an agent-based computational model with spot-futures cross-market structure which coincides with the main characteristics of the Chinese stock market and the CSI index futures market. This model can be useful in analyzing asset pricing, information transmission and risk spreading between the futures market and the spot market, as well as the implementation of hedging strategies.

## 2. The Model

In this model, there exists a stock market where several stocks are traded and a stock index futures market where one stock index future is available. For each market, there are three types of investor, namely informed traders, uninformed traders and noise traders. What is more important is that there exists a kind of spot-futures arbitrageurs who build or close positions in both markets. Arbitrageurs' behaviors lead to the co-movement between the two markets. For each type of traders, the investment demands are endogenously determined, and are subject to constraints from wealth, risk management level and trading mechanisms in the market, which is largely in line with the present trading mechanisms of the Chinese stock market and the stock index futures market.

### 2.1. Assets

In the artificial markets, we assume that there are five stocks and one stock index futures whose underlying asset is the stock index, which is constructed based on these five stocks' prices.

*1) Stocks' common value*

The evolution of the stock's common value $v^*_{i,t}$ is given by

$$v^*_{i,t+1} = (1+\phi_i+\sigma_{i,\varepsilon}\varepsilon_{t+1}) v^*_{i,t} \quad (1)$$

where $\phi_i$ is the dividend growth rate of stock $i$, namely the random walk drift of the common value. The simulation time $t$ corresponds with some time interval in the real world, say, 5 seconds. We set the growth rate to be zero due to the extremely short time interval, which means $\phi_i = 0$. We assume $\varepsilon_t \in N(0,1)$. $\sigma_{t,\varepsilon} > 0$ stands for the standard deviation in this diffusion process. Parameters of the five stocks are set as Table 1.

*2) Stock Index*

The unit of the stock index is called "point", and the base period value of the index is 3000 points. The stock index is calculated using the weighted composite price index method, which is showed as follows

$$I_t = M_t / M_0 \times 3000 = (\Sigma p_{i,t} S_{i,t}) / M_0 \times 3000 \quad (2)$$

where $M_t$ is the market value of index stocks at period $t$, which is the sum of current stock price $p_{i,t}$ multiplied by the number of shares outstanding $S_{i,t}$, and $M_0$ is that of base period.

*3) Index futures' common value*

The common value of stock index futures will be calculated according to the futures' theoretical value, that is, $v_{F,t} = I_t (1+r)^{T-d+1}$, where $I_t$ is the current index, $T$ is the expiration date, and $d$ is the days that the stock index future has been listed.

**Table 1. Parameters of the five stocks**

| Stock No. | Initial Value | S.D. of Disturbance | Stock Shares(100 million) |
|---|---|---|---|
| 1 | 10 | 0.0007 | 50 |
| 2 | 20 | 0.0007 | 40 |
| 3 | 30 | 0.0003 | 60 |
| 4 | 40 | 0.0003 | 30 |
| 5 | 50 | 0.0005 | 50 |

### 2.2. Markets

- There are two markets, a stock market which contains five stocks and a stock index futures market which contains one index futures. Both markets adopt continuous double auction trading mechanism and the $T$+0 rule, considering the need of high-frequency trading. Investors can place limit orders or market orders. Buy limit orders whose prices are equal to or higher than the

- best ask price and sell limit orders whose prices are equal to or lower than best bid price are treated as market orders and executed instantly. According to the rules of the Chinese Financial Futures Exchange, limit orders have certain life time, that is, unexecuted limit orders will be cleared at the end of the day.
- The simulation time $t$ is similar to a five-second interval in real world, and there may be several transactions or no transaction during the time.
- Short selling is not allowed in the stock market, but is allowed in the futures market. Thus arbitrageurs can only engage in positive arbitrage, that is, they can only buy stocks and sell futures.
- The stock index future market claims margin, and settles balance of each margin account after the market close every day. Once the investor's equity is less than the minimum margin requirement for holding positions, his positions will be forced to close one by one the next day, until his equity is no less than the required minimum margin.
- The transaction price of the market $p_{i,t}$ is the average price of the several transactions during $t$, and if no transaction is done, the price is equal to that of last time, namely $p_{i,t} = p_{i,t-1}$.
- There exist no transaction costs for both stocks and futures.
- If the wealth of an investor is smaller than a certain amount, he will be identified as bankrupt and exit the market. Another investor who is the same with the former one in investor type, initial wealth and stocks or futures investment will enter the market.

## 2.3. Traders structure and their behaviors

There are seven kinds of investors in the markets. Investors that trade only stocks include informed traders, uninformed traders and noise traders. For each of them, they are designed to randomly choose one stock and invest on it permanently. Also there are three similar types of traders that trade only in the futures market. The spot-futures arbitrageurs trade in both stock market and futures market.

The spot-futures arbitragers real-timely watch the relationship between the stock index and the price of the stock index futures. Once the futures price is higher than the upper boundary of the no-arbitrage range and reaches the arbitrageur's expected profit point, he will take positions and buy stock portfolios and sell futures in the same time. He will close the position (sell stock portfolios and buy futures) once the futures price falls back to the arbitrage caps, otherwise he will hold the positions to due. The arbitrageurs seek the immediate execution of their orders, thus they only place market orders. To realize risk-free arbitrage, the spot-futures arbitrageurs allocate their wealth between stock portfolios and futures, and keep the ratio of the margin for the whole assets within a specific range keeping the account safe.

The heterogeneous expectations and order submission size for six kinds of single-market traders are described as follows.

*1) Trader's expectation on asset price*

Since the expectations of the three types of investors who trade only in the stock market are similar to those who trade only in the futures market, we talk about them together and remove the subscript that stands for the asset number for simplicity. Assume that all the investors know the current common value of a stock, while there is divergence in the futures common value.

a) Informed trader $i$ accurately know the common value of the stock in the following $\tau$ periods when he enters the market, and the expected price is

$$\hat{p}^i_{t+\tau} = v_{t+\tau} \qquad (3)$$

However, as for informed traders in futures market, they cannot accurately know $v_{t+\tau}$ as they are not able to predict $I_{t+\tau}$ accurately. Here we assume that they learn $\hat{I}_{t+\tau}$ through stock's common value, their expected price $\hat{p}^i_{t+\tau}$ will be

$$\hat{p}^i_{t+\tau} = \hat{I}_{t+\tau}(1+r)^{T-d+1} = (\Sigma v_{i,t+\tau} S_{i,t})/M_0 \times 3000 \times (1+r)^{T-d+1} \qquad (4)$$

b) Uninformed traders cannot know the future common value $v_{t+\tau}$, but they know current common value $v_t$. They obtain their expected prices by mixing three sources, including the current common value $v_t$, $\tau$-period average transaction prices $\bar{p}_\tau$ and current midpoint of bid and ask prices $p_m$. The expected price is given by

$$\hat{p}^i_{t+\tau} = (a^i v_t + b^i \bar{p}_\tau + c^i p_m) / (a^i + b^i + c^i) \qquad (5)$$

c) The expected price of nosie trader $i$ is randomly chosen within the five levels from the bid and ask prices in the order book, which is given by

$$\hat{p}^i_{t+\tau} = bid_5 + rand^i_t \times (ask_5 - bid_5) \qquad (6)$$

*2) Order size*

Given price $p$, an investor's optimal positions depend on his utility function. Following the demand determined in [11], we assume that investors are absolutely risk averse, namely they make their investment decision by maximizing CARA utility function, and their optimal positions is

$$\pi^i(p) = \ln(\hat{p}^i_{t+\tau} / p) / (\alpha^i V^i_t p) \qquad (7)$$

where $\alpha^i$ is the absolute risk averse coefficient of investor $i$, $V^i_t$ is the variance of expected return of investor $i$, $\hat{p}^i_{t+\tau}$ is expected price at period $t+\tau^i$, which is different for different types of investors, $p$ is the order submission price. If the demand quantity $\pi^i(p)$ is larger (smaller) than the investor's current position, then the investor buys (sells). $V^i_t$ is estimated by the variance of past returns:

$$V^i_t = \Sigma^\tau_{j=1}(r_{t-j} - \bar{r}^i_t)^2 / \tau^i \quad (8)$$

$$\bar{r}^i_t = \Sigma^\tau_{j=1} r_{t-j} / \tau^i = \Sigma^\tau_{j=1} \ln(p_{t-j}/p_{t-j-1}) / \tau^i \quad (9)$$

### 2.4. Parameter Settings

The evolution of common value of the five stocks has been described. Here we only specify parameter settings on the markets and the investors. Most of these parameters are consistent with the market structure and trading rules in the Shanghai Stock Exchange and the China Financial Futures Exchange.

Every simulation will run 54758 steps, corresponding to 19 days. Namely, each day contains 2882 steps and each step stands for five seconds. The risk-free interest rate in our model is 8%, which is close to the annual lending interest rates in China. The order book is cleared every day in both stock market and futures market. The minimum order size required in the stock market is 100 shares as in the real market. The tick size is 0.01 Chinese Yuan for stocks and 0.2 Chinese Yuan for the futures. The minimum margin rate is 18%, which is the margin rate required by brokers in real trading. The multiplier of futures contract in our model is 300 Chinese Yuan per point, which is also consistent with the IF1009 contract.

For the three types of investors in the stock market, each investor permanently invests on one stock which is randomly selected before the program begins. The initial positions are allocated randomly between the range [300, 1500], and the initial cash is equal to the initial values of all his stocks. For the three types of investors in the futures market, each one's initial wealth is 3 million, and for each spot-futures arbitrageur it is 10 million. The orders submission interval for the six kinds of single-market investors obeys an exponential distribution, that is, the number of orders during one period follows the Poisson distribution. At the same time, for these six types of single-market investors, the order submission interval is equal to the order life. If they find limit orders submitted before are still unexecuted or not fully executed when they reenter the market, they will cancel the former orders first and then place new orders. For uninformed traders, the variables $a^i$, $b^i$ and $c^i$ that represent the weights when forecasting prices are randomly chosen from [0, 1]. The capital safety ratio for futures investors, which is the wealth share for futures investment, is no more than 60%. As for arbitrageurs' ex ante expected profits for every futures contract, which should be enough to compensate the execution costs when trading in both two markets, we set them randomly in the range [10, 20], which are equivalent to [50, 100] index futures' tick sizes. Each investor's wealth will be updated at the end of every period. If one's wealth is less than the capital to buy 100 shares of his invested stock, he goes bankruptcy and exits the market, while a new investor with the same initial settings as the former one will enter the market. So it is in the futures market when a futures investor can't afford one index futures contract.

## 3. The simulation results and the statistical properties

### 3.1. The first results

We employ basis, best bid-ask spread, volatility clustering and autocorrelation of returns as well as other key indicators of market features to compare the agent-based computational cross-market model with the 5-second high-frequency data of IF1009 contracts from August 20th, 2010 to September 15th, 2010. Among these indicators, volatility clustering and autocorrelation of returns are common stylized facts in financial markets that adopt continuous double auction trading mechanism [12]. At the same time, they are fundamental for the calibration of agent-based computational models.

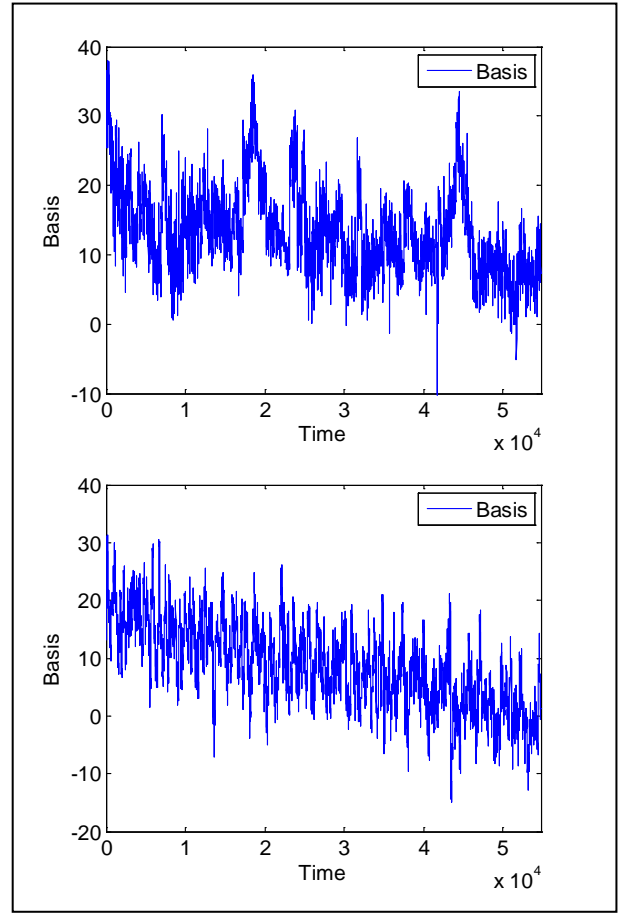

**Figure 1. Time series of spot-futures basis (Upper Panel: basis between IF1009 and CSI 300; Lower Panel: basis between simulated stock index and simulated index futures).**

Fig. 1 shows the spot-futures basis series of the simulated data and the real data. The spot-futures basis is defined as

$$\rho = F - S \quad (10)$$

We can find that the two groups of basis have similar downward trends which reflect the convergence of stock index and index futures as the maturity date is approaching. Further, the distributions of spot-futures basis are shown in Fig. 2. Similar to the basis between IF1009 and CSI 300, the basis from our simulated model can also be well fitted by the general extreme value distribution.

The distributions of bid-ask spreads are shown in Fig. 3. Both the spreads of IF1009 and that of our simulated futures decreases similarly and the magnitudes of each statistics are roughly identical.

The distributions of logarithmic returns are shown in Fig. 4, where both IF1009 and the simulated futures show the same properties of leptokurtic and fat tails.

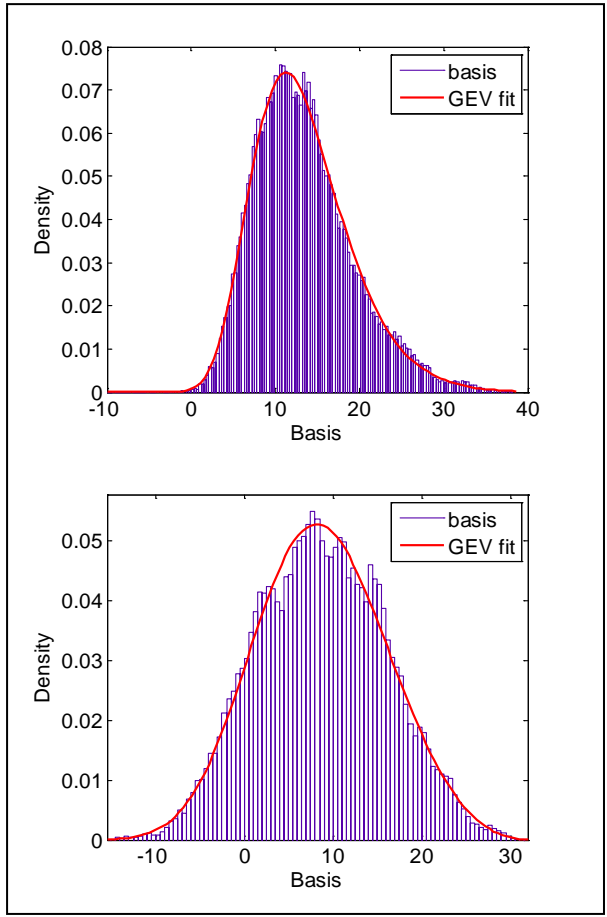

**Figure 2. Distribution of spot-futures basis (Upper panel: distribution of basis between IF1009 and CSI 300; Lower panel: distribution of basis between simulated stock index and simulated index futures).**

Fig. 5 calibrates the long-term memory property of real IF1009 return and the simulated index futures return. Autocorrelation function plots of absolute returns can be used to verify long-term memory. We can also see that Both IF1009 and our simulated futures have similar long-term memory.

Martens suggested that GARCH model can better describe the volatility of stock index futures [13]. To examine the volatility clustering of real data and simulation data, we conduct the test using GARCH(1,1) model, which is shown as follows

$$r_t = ar_{t-1} + br_{t-2} + \varepsilon_t \quad (11)$$

$$\sigma_t^2 = c + \alpha \varepsilon_{t-1}^2 + \beta \sigma_{t-1}^2 \quad (12)$$

Where $\beta$ is coefficient of GARCH and represents the degree of volatility clustering. A larger $\beta$ means higher volatility clustering. From table 2, we can find that the return series of both IF1009 and simulated futures have similar volatility clustering. They all follows an AR(2)-GARCH(1,1) process and the values of $\beta$ are 0.81 and 0.90 respectively.

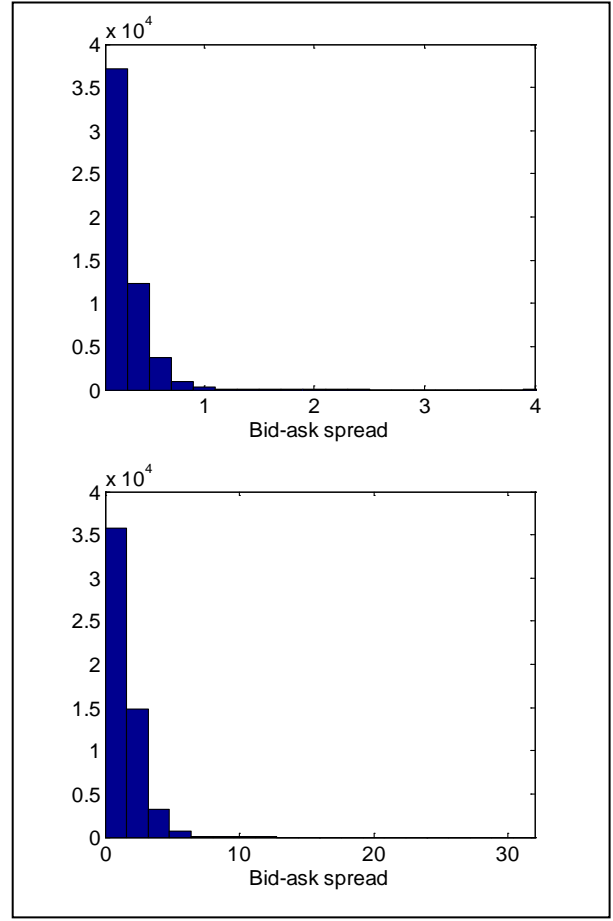

**Figure 3. Distribution of bid-ask spreads (Upper panel: IF1009; Lower panel: simulated index futures).**

After comparing the spot-futures basis, the bid-ask spread, volatility clustering and autocorrelation of absolute return, we figure that our agent-based computational model can reproduce the key features of the real spot market and the stock index futures market in China.

**Table 2. AR(2)-GARCH(1,1) parameter values estimated for the return series of IF1009 and simulated futures**

| Series | a | b | c | α | β |
|---|---|---|---|---|---|
| IF 1009 | -0.047506 (0.004510)* | -0.021214 (0.004602)* | 1.41E-09 (3.40E-11)* | 0.154731 (0.002074)* | 0.816699 (0.002024)* |
| Simulated Future | -0.525263 (0.004278)* | -0.234303 (0.004271)* | 1.37E-09 (7.66E-11)* | 0.082618 (0.002283)* | 0.900256 (0.002667)* |

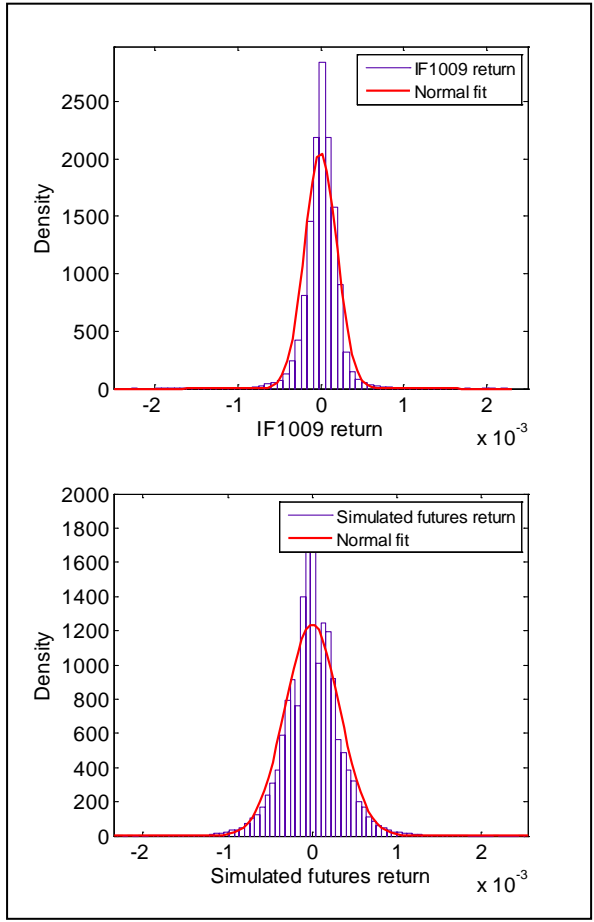

**Figure 4. Distribution of logarithmic returns (Upper panel: IF1009; Lower panel: simulated index futures).**

### 3.2. More experiments

Further, we carry out more computational experiments to clarify whether the simulation results are stable and whether the parameter settings of the five stocks would affect the outcomes of the model.

There are three parameters for stock common value setting: the initial value $v^*_{i,0}$, the standard deviation of disturbance $\sigma_{i,\varepsilon}$ and stock shares in the market. It is easy to understand that the initial value will have no obvious impact on the simulation results. In addition, stock shares in our model are only used to calculate the weights for stock index, so what may affect the outcomes of the model, if any, are disturbances' standard deviations, not the weights, intrinsically. In view of this, we design two extra simulations for different standard deviations of disturbances as in Table 3.

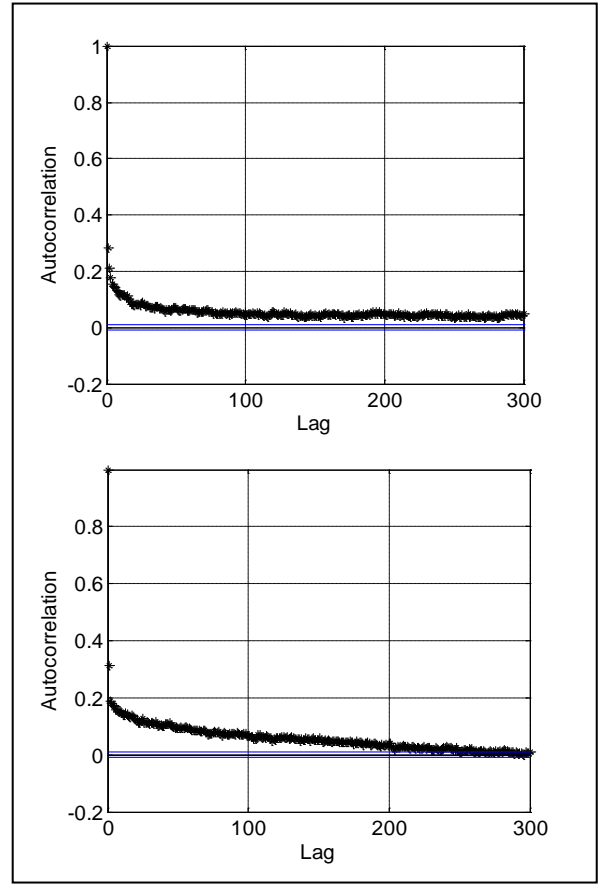

**Figure 5. Autocorrelation plots of absolute returns (Upper panel: IF1009; Lower panel: simulated index futures).**

**Table 3. Two simulations for different standard deviations of disturbances**

| Stock No. | Initial Value | S.D. of Disturbance | |
|---|---|---|---|
| | | Simulation 1 | Simulation 2 |
| 1 | 10 | 0.0005 | 0.0008 |
| 2 | 20 | 0.0005 | 0.0007 |
| 3 | 30 | 0.0007 | 0.0006 |
| 4 | 40 | 0.0007 | 0.0005 |
| 5 | 50 | 0.0003 | 0.0004 |

Figs. 6-10 show the results of the comparative experiments. We can find that for all statistics, namely spot-futures basis, bid-ask spread, logarithmic return and its autocorrelation, the two simulations show similar statistical properties as the real market and the first results in the preceding subsection. Thus, we can conclude that the results of this cross-market model are stable and this model can reproduce the key features of the real spot market and the stock index futures market in China.

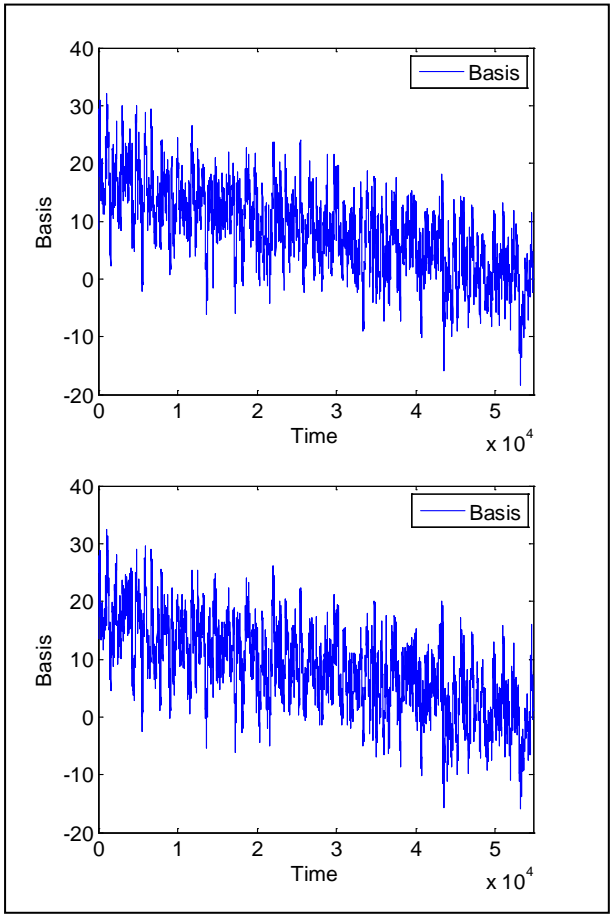

**Figure 6. Time series of spot-futures basis (Upper Panel: Simulation 1; Lower Panel: Simulation 2).**

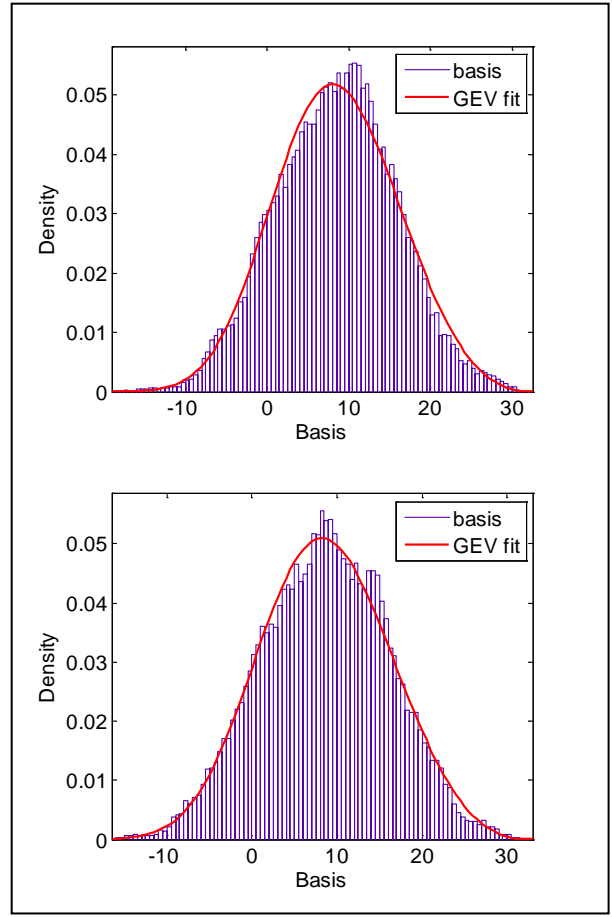

**Figure 7. Distribution of spot-futures basis (Upper Panel: Simulation 1; Lower Panel: Simulation 2).**

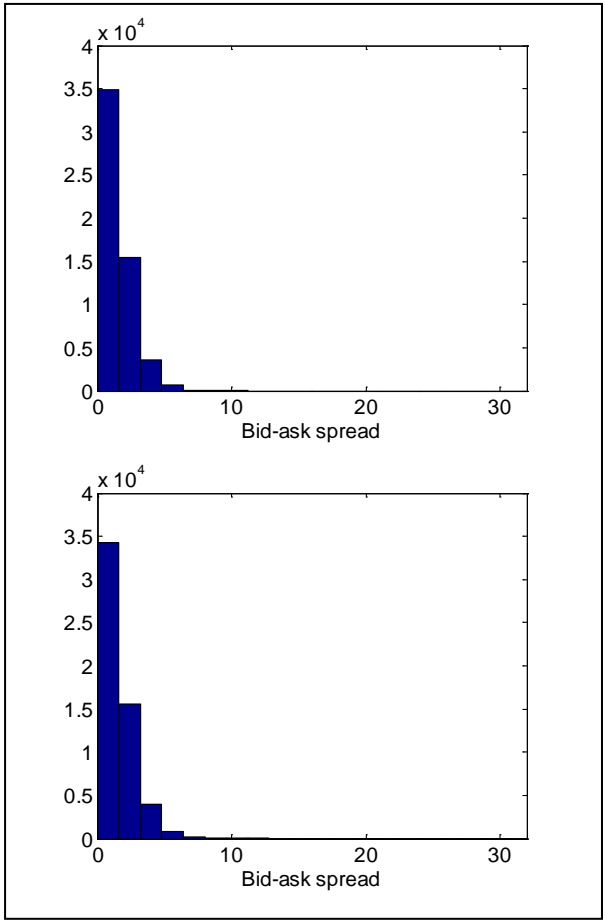

**Figure 8. Distribution of bid-ask spreads (Upper Panel: Simulation 1; Lower Panel: Simulation 2).**

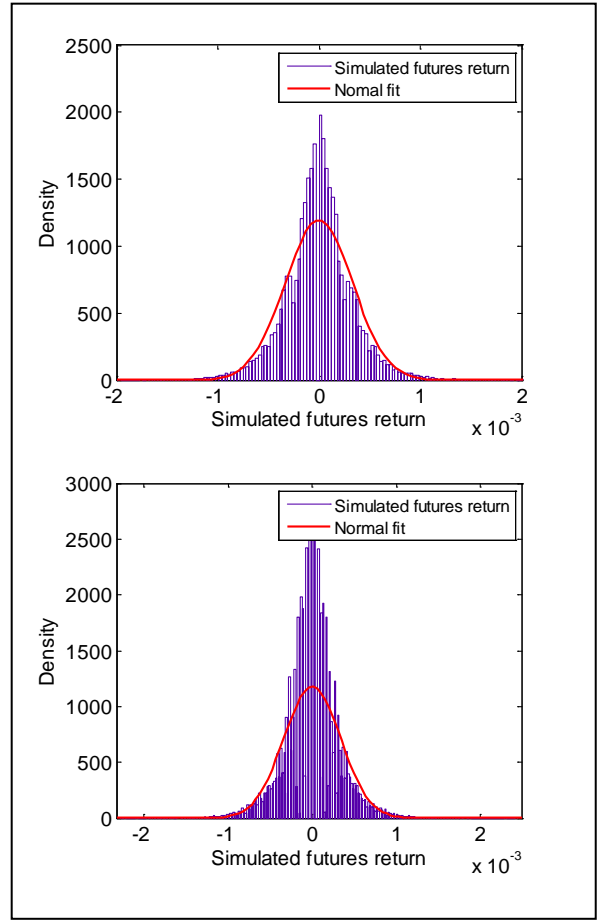

**Figure 9. Distribution of logarithmic returns (Upper Panel: Simulation 1; Lower Panel: Simulation 2).**

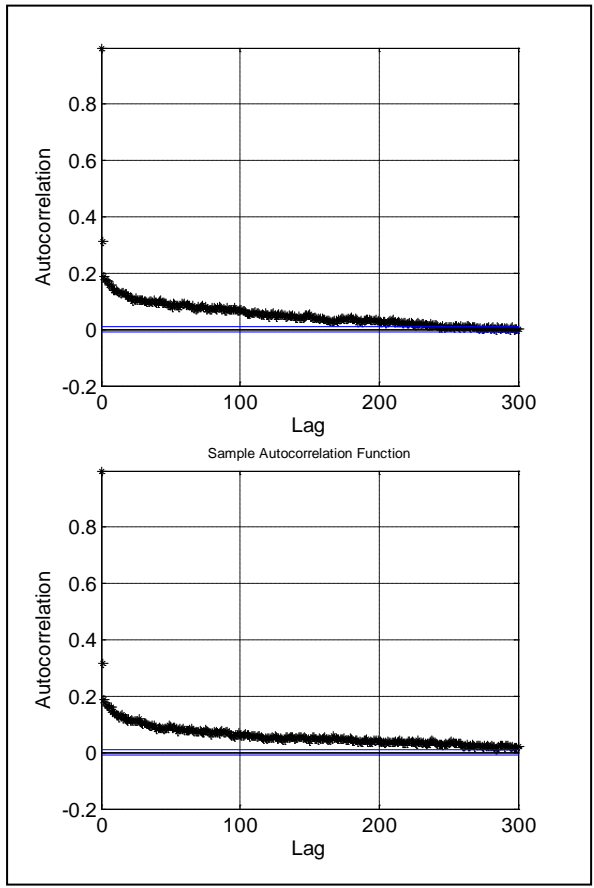

**Figure 10. Autocorrelation plots of absolute returns (Upper Panel: Simulation 1; Lower Panel: Simulation 2).**

## 4. Concluding remarks

According to the characteristics of CSI 300 index futures market and Shanghai stock market, such as continuous double auction trading mechanism, investor types and investment strategies, we build up an agent-based computational model with spot-futures cross-market structure. By analyzing the key properties including spot-futures basis distribution, bid-ask spread distribution, volatility clustering and autocorrelation of absolute returns, and then comparing them with those of real market data, we show that our model can successfully reproduce the Chinese financial markets' features. Thus, the model can be applied to studies such as inherent risk diffusion, critical factors of cross-market risk conduction, trading mechanisms, reasonability of investor structure, and execution risk of trading strategies.

Furthermore, what is the impact of the introduction of stock options and application of related hedging strategies on Chinese stock market? It will be a useful extension to include stock options in our current model to explore the equity markets from a systemic view.

## Conflict of Interests

The authors declare that there is no conflict of interests regarding the publication of this paper.

## Acknowledgment

We thank Editor Pankaj Gupta and anonymous referees for useful suggestions that helped us improve the paper substantially. This research is partly supported by NSFC (No. 71320107003, 71131007, 71271145), China Program for New Century Excellent Talents (NCET-07-0605), Program for Changjiang Scholars and Innovative Research Team (IRT1028).